%% file: main.tex
\documentclass[
preprint,
superscriptaddress,
amsmath,
amssymb,
aps,
prc
]{revtex4-2}

\usepackage{graphicx}
\usepackage{dcolumn}
\usepackage{bm}
\usepackage{hyperref}
\usepackage{float}
\usepackage{subcaption}

\begin{document}

\title{Formation of Quark-Gluon Plasma droplets under Ultra-Relativistic Heavy Ion Collisions in the presence of magnetic fields with a modified hadronic medium}

\author{Ananya Sharma}
\affiliation{Department of Physics, Kirori Mal College, University of Delhi, Delhi, India}
\affiliation{International Centre for Theoretical Sciences, Tata Institute of Fundamental Research, Bengaluru, India}

\author{Divya Rathee}
\affiliation{Department of Physics, Kirori Mal College, University of Delhi, Delhi, India}

\author{Agam K. Jha}
\affiliation{Department of Physics, Kirori Mal College, University of Delhi, Delhi, India}


\begin{abstract}
\input{abstract}





 
\end{abstract}
\maketitle

\newpage

\section{Introduction}
\input{introduction}

\section{Formalism}
\input{formalism}

\section{Results}
\input{results}

\section{Conclusions}

We conclude from our study that kaons, along with pions, in the medium of the QGP droplet result in a higher free energy value, hence implying a more stable QGP droplet with a larger radius. Further, the incorporation of effective mass, chemical potential and an external magnetic field stabilises the system and also increases its stability. $\mu$ and B have opposing effects on the system. As $\mu$ is increased, the stability decreases, whereas when the magnetic field is increased, the stability of the system increases. These opposing effects of both these quantities point towards finding finite-sized and stable QGP droplets. The phase diagram shows us the most stable regions in the parameter space of the magnetic field and the chemical potential lie in high magnitudes of magnetic field and low values of the chemical potential. The heat map also shows the configurations wherein a QGP droplet cannot be formed, those being the regions where chemical potential is high and magnetic fields are low. This region shifts to the right as chemical potential is increased, implying that QGP formation requires a higher magnitude of magnetic field as the chemical potential increases.\\

The phase transition in this system is not of the first order, as there is no discontinuity observed at the critical transition temperature. However, an analysis of the heat capacity shows that it follows a downward curve after a certain temperature is achieved, implying a phase transition. Hence, it may be said that the phase transition is weakly first-ordered. The results are in agreement with lattice QCD calculations.\\

\begin{acknowledgments}
The authors gratefully acknowledge support from the Department of Physics, Kirori Mal College, University of Delhi for providing the academic environment and institutional support necessary for carrying out this research. We also thank the faculty members of the department for their discussions and encouragement throughout the course of this work.  
\end{acknowledgments}

\bibliographystyle{apsrev4-2}
\bibliography{references}

\end{document}

%% file: abstract.tex
We build upon the previous studies conducted in this domain using the density of states of various components of the quark-gluon plasma(QGP) modelled after the Thomas-Fermi and Bethe models. This study provides an analysis of the stability of the quark-gluon plasma droplet by the inclusion and variation of chemical potential and magnetic field. An effective mass term is included to account for both the rest mass and thermodynamic mass terms of the quarks and gluon. We also expand on the inclusion of a different meson, kaon, into the medium of the droplet. The analysis on these quantities is done by calculating the free energy of the system and examining it as a function of droplet radius, chemical potential and magnetic field to find conditions for stable droplet formation. We discover that a kaonic medium, along with a pre-existing pionic medium, provides greater stability to the system. At the same time, it is realised that the parameters of chemical potential and magnetic field, due to their opposing effect, help in containing the droplet. Further calculations on entropy and heat capacity yield information about the nature of the phase transition of the system. Our analysis and results are in accordance with lattice QCD simulations.\\

\vspace{2cm}
\textit{Keywords}: Quark-Gluon Plasma, Quantum Chromodynamics, Ultra-Relativistic Heavy Ion Collisions, Quark-Hadron Phase Transitions\\
PACS Numbers: 12.38.Mh, 25.75.-q, and 24.85.+p

%% file: introduction.tex
 The physics of quark-gluon plasma is too complicated to be studied directly through quantum chromodynamics(QCD), a true theory of strong interaction; hence, we continue to adopt a thermo-statistical approach \cite{statisticalmodel1, statisticalmodel2} in trying to understand the quark-gluon plasma using the Thomas-Fermi and Bethe Models \cite{thomasfermi, bethe}. The models have been adjusted to take into account the hydrodynamical and high-temperature aspects of QGP. Experimental research on QGP today is done through Ultra Relativistic Heavy Ion Collisions(URHIC) at LHC, CERN and RHIC, BNL. Incorporating other parameters into this model is motivated by the conditions in these experiments as well as to make the QGP droplet more stable. Upon collisions, the QGP droplet consists of different flavours of quarks and gluons which exist in a mesonic medium. Previous studies only assume a pionic medium for the droplet \cite{magnetism, prc04}; however, experiments point towards the presence of kaons as well\cite{kaon}. Hence, we are motivated to study the system in a pion-kaon medium for the QGP. The relativistic collisions of particles also cause a change in the particles' rest mass, which has a thermodynamic dependence. Hence, rest mass and the thermodynamic dependence need to be combined to give an effective mass. We also study the effect of an external magnetic field on the system. The particle collider experiments do not have head-on collisions; as a result, the QGP droplet so formed is surrounded by an external magnetic field. It is hypothesised that QGP existed shortly after the Big Bang. The large-scale magnetic fields in today's galaxies may be a result of the primordial magnetic fields of the early universe\cite{primordialB}. Hence, this also motivates us to include an external magnetic field in our calculations to form a truer picture of the QGP. The chemical potential is another important QCD parameter along with temperature. It is related to the number of particles in the QGP droplet. Hence, by varying the chemical potential, we study its effect on the system. \\

We include these parameters in our calculations and study their effect on the total free energy of the system. We compute the free energy and look at it as a function of various parameters and study the effects of these parameters. Through free energy, we also calculate the entropy and heat capacity of the system to understand the nature of the phase transition in this altered system. 

%% file: formalism.tex
\subsection{Free energy and other expressions:}

In previous studies \cite{prc04, magnetism}, which included a magnetic field, chemical potential and thermodynamic mass dependence, the thermal potential \cite{thermalpot, dynamicmass} was modified and defined as follows: 

\begin{equation}
    (V_{conf})_{q,g} = \frac{1}{2k} \left(\left(\gamma_{q,g}\right) g^2(k)\, T^2 -M_{{d}_{g,q}}\right)
\end{equation}

where $M_{d_{g,q}}$ is the thermodynamic mass term dependent on temperature \cite{magnetism, dynamicmass}, k is the four-momentum, $g(k)$ is the QCD coupling constant, T is the temperature, and $\gamma_{q,g}$ are the phenomenological flow parameters which account for the hydrodynamical nature of the QGP droplet\cite{thermalpot,dynamicmass}. This is the confining potential for quarks and gluons, which accounts for their bulk properties. Since the magnetic field does no work in the confinement of these particles, it does not show up in the potential term. The other parameters are defined as: 
\begin{equation}
    M_{d_{g,q}} = \gamma_{q,g} T^2 \left(\frac{1}{\ln (1+T/T_c)}\right)^2 \left(\frac{16\pi^2}{11-\frac{2}{3} N_{f_{q,g}}}\right)
\end{equation}

\begin{equation}
    g^2(k) = \frac{4}{3} \frac{12\pi}{27} \frac{1}{\ln(1+k^2/\Lambda^2)}
\end{equation}

where $T_c$ is the critical transition temperature equal to 175 MeV, $N_{f_{q,g}}$ is the flavour number, with values of 3 for quarks, as this study considered only up, down and strange quarks in the system, and 0 for gluons\cite{dynamicmass}. $\Lambda$ is the QCD parameter, equal to 150 MeV. We can see in this expression, eqn. (2), that mass is a function of temperature. This expression does not take into account the rest masses of the particles. The only distinction between particles is due to their flavours.
The density of states is defined as\cite{thomasfermi, bethe}: 
\begin{equation}
    \rho_{q,g}(k) = \frac{v}{\pi^2} \left[\left(V_{conf}(k)\right)^2 \left(-\frac{dV_{conf}}{dk}\right)\right]
\end{equation}
where $v$ is the volume of the droplet, which is simply $\frac{4}{3} \pi R^3$. The free energy for quarks $(-,+)$ and gluons $(+,-)$ is therefore defined as:

\begin{equation}
    F_{q,g} = \mp T\, g_{q,g} \int dk\, \rho_{q,g}(k) \ln\left(1\pm e^{-\left(\sqrt{M_{d}^2+k^2+B}-\mu\right)/T}\right)
\end{equation}
Here, $\rho_{q,g}(k)$ is the density of states of quarks or gluons, which represents the number of states having momentum between $k$ and $k+dk$, and $g_{q,g}$ is the degeneracy factor whose value is 6 for quarks and 8 for gluons. The previous models also assume the surface surrounding the droplet to be a Weyl surface\cite{weylsurface, magnetism, statisticalmodel2}. Its free energy is: 

\begin{equation}
    F_{surface} = \frac{1}{4}\, R^2\,\left(T^3\, \gamma\, +\, B\, *\,T\right)
\end{equation}
where $\gamma$ is the inverse r.m.s value of the flow parameter of quarks and gluons and equals: 
\begin{equation}
    \gamma = \sqrt{\frac{2}{\gamma_g^2}+\frac{2}{\gamma_g^2}}
\end{equation}
The free energy of the pion is defined as\cite{pionFE}: 
\begin{equation}
    F_\pi = - \frac{3Tv}{2\pi^2} \int dk\, k^2\, \ln\left(1-e^{-\left(\sqrt{M_\pi^2+k^2+B}-\mu\right)/T}\right)
\end{equation}

It should be noted that the mesons in the medium still have only a rest mass term, not an effective mass term. This is due to the relation of pions and kaons with chiral symmetry \cite{magnetism}. At the same time, the inclusion of a magnetic field reverses the signs of the free energy of the quarks and gluons; hence, in the final expressions, the signs are reversed when there is an external magnetic field present\cite{magnetism}.  The above definite integrals are evaluated with regard to a specific value of $k$, termed $k_{min}$, which is the low-energy cut-off of the system. The expression is integrated over $k_{min}$ to $5k_{min}$ , where the upper limit is chosen such that the integral of free energy has saturated by those values\cite{statisticalmodel2}. 
\begin{equation}
    k_{min} = \left(\gamma_{q,g}\, \frac{4}{3} \frac{12\pi}{27}\, T^2\, \frac{\Lambda^2}{2}\right)^{1/4}
\end{equation}

An important feature of the discussion surrounding QGP is the nature of the phase transition. It is widely considered that the phase transition is of the first order \cite{o1phasetrans}. This is determined by calculating the entropy of the system. It is calculated by: 
\begin{equation}
    S = - \frac{\partial F}{\partial T}
\end{equation}
If we are to conclude that the phase transition is of the first order, the entropy should show a discontinuous nature at the critical transition temperature. The previous studies reflect the above\cite{magnetism}. We shall investigate if the same is true when new parameters are included in this study. \\

Calculating the heat capacity also yields information about phase transitions. If there is a sharp decrease in heat capacity, it indicates a phase transition in the system. Heat Capacity is calculated as: 
\begin{equation}
    C_v = T \frac{\partial S}{\partial T} = -T \frac{\partial^2 F}{\partial T^2}
\end{equation}

\subsection{New Parameters}
\subsubsection{\textbf{Effective Mass}}
As mentioned before, eqn. (2) for $M_d$ does not take into account the rest mass of the particles. Hence, we seek an expression that combines the dynamic mass term with the rest mass. We call this the Effective Mass and replace the dynamic mass term in the above expressions with $M_{eff}$ \cite{Meff1, Meff2}. 
\begin{equation}
    M_{eff}^2 = m_0^2 + \sqrt{2} m_0m_d + m_d^2
\end{equation}
Where $m_0$ is the rest mass and $m_d^2=M_d$. Since QGP is produced as a result of ultra-relativistic collisions, we have made sure that any new formulas, such as the above, are Lorentz invariant. 
Hence, the change in potential and free energy of quarks and gluons is as follows: 
\begin{equation}
    V_{conf}(k) = \frac{1}{2k} \left(\gamma_{q,g}\, g^2(k)\,T^2-M_{eff}^2\right)
\end{equation}
\begin{equation}
    F_{q,g} = \mp T\, g_{q,g} \int dk\, \rho_{q,g}(k) \ln\left(1\pm e^{-\left(\sqrt{M_{eff}^2+k^2+B}-\mu\right)/T}\right)
\end{equation}

\subsubsection{\textbf{Kaonic Medium}}
Another important addition in this study is that of kaons. Kaons, too, are mesons like pions made up of strange quarks with a degeneracy of 4. The debris of particle collision contains both pions and kaons\cite{kaon}; however, previous studies have only considered the system to be in a pionic medium\cite{magnetism,prc04}. In our study, we want to take the system under analysis closer to the true state of QGP formation; hence, we want to investigate the effect on the system as a result of the inclusion of kaons. The kaonic free energy is constructed analogously to pionic free energy, which is: 
\begin{equation}
    F_K = - \frac{4Tv}{2\pi^2} \int dk\, k^2\, \ln\left(1-e^{-\left(\sqrt{M_K^2+k^2+B}-\mu\right)/T}\right)
\end{equation}
We study the change caused in the system as a consequence of the above addition. We do so by comparing it to our analysis in a system without kaons. The inclusion of kaons introduces the effect of strangeness into the hadronic medium, thereby modifying the thermodynamic behaviour of the QGP droplet. Since kaons are heavier than pions, their contribution to the hadronic free energy is expected to influence the stability of the droplet differently. By comparing the free energy profiles of pionic and pionic-kaonic media, we analysed how the presence of kaons affects the formation, stability, and phase transition characteristics of QGP droplets under URHIC.

\subsubsection{\textbf{Chemical Potential and Magnetic Field}}
We also study the chemical potential $\mu$\cite{chempotential}, the analysis of which has been lacking in previous studies. The chemical potential is an important QCD parameter along with temperature and is integral to the QCD phase diagram. It represents the number of particles in the system. We vary the value of the chemical potential between $50-500$ MeV and study, again, its stability on the system. This range is selected in accordance with lattice QCD calculations wherein the system is stable.\cite{chempotential}\\

We also expand on the range of magnetic field included in previous analyses. Previous studies provide analysis on the range between $15m_{\pi}^2$ and $30m_\pi^2$\cite{magnetism}. These values were in accordance with LHC experiments \cite{1LHC.B, 2LHC.B, 3LHC.B}. Here, in our extended analysis, we study the magnetic field from $0m_\pi^2$ to $45m_\pi^2$. This range is selected in accordance with lattice QCD calculations \cite{B.range}. It is important to note here that for any given $x*m_\pi^2$ equals $eB$, where $e=0.303$, the electric charge in the natural system of units, acts as the coupling constant, and $m_\pi$ is the mass of the pion.
\begin{equation}
    B = x * \frac{m_\pi^2}{e}
\end{equation}

The total free energy, then, becomes: 
\begin{equation}
    F_{total} = F_u\, +\, F_d\, +\, F_s\,+\,F_g\, +\, F_{\pi} +\, F_{K}\, +\, F_{surface}
\end{equation}

Free energy reflects the stability of the system, making it integral to our analysis. A free energy graph with a non-infinite peak that ultimately goes to negative infinity reflects a stable QGP droplet. Also, the higher the value of the non-negative peak, the more stable the system is \cite{F1, F2, F3, F4, F5}. Our analysis of various parameters is not just limited to studying free energy as a function of the radius of the droplet. In this study, we have looked at the free energy as a function of the thermodynamic parameters: chemical potential and magnetism. We have also analysed the relationship between chemical potential and magnetism using phase diagrams to identify the most stable configurations of the QGP within their respective ranges. \\

%% file: results.tex
We shall present all our analyses by comparing cases for pionic and pionic + kaonic media, including the parameters mentioned above as we go. We shall also present graphs that differ from those in previous studies to support our analysis.

\subsection{Effect of Hadronic Medium on the Stability of QGP Droplets.}
Fig. 1 shows the variation of the free energy with droplet radius of all components of the QGP droplet in the pionic and pionic + kaonic medium.
\begin{figure}[b]
    \centering
    \includegraphics[width=0.65\linewidth]{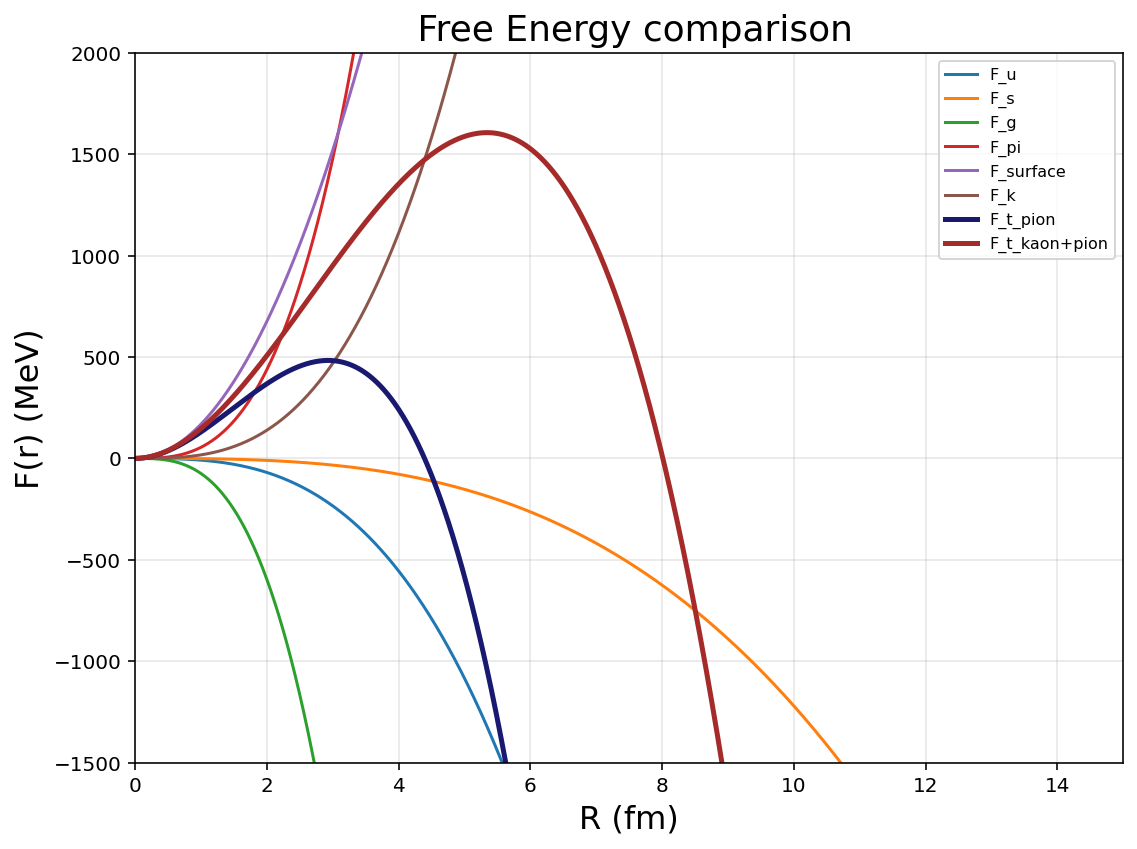}
    \caption{Variation of free energy in kaonic + pionic vs only pionic medium in the original system along with individual components}
    \label{fig:placeholder}
\end{figure}

From the graphs, the total free-energy curve for the pionic medium produces a smaller slope than the kaonic-pionic medium.

We can see clearly from the free energy analysis that the total free energy peak is increased when kaons are included in the system. It indicates that the droplet is more stable in kaonic+pionic than in a pure pionic medium, as the higher the peak of free energy, the more stable the QGP droplet.

\subsection{System in Magnetic Field and non-zero Chemical Potential}
Starting from the parameters included in the previous studies, we include chemical potential $\mu=400$ MeV, a dynamic mass term $M_d$, and a magnetic field, B = 933345 MeV$^2$. Fig. 2 clearly shows an increase in the free energy when kaons are included in the droplet medium. Hence, a kaonic-pionic medium is considerably more stable with these parameters too.

\begin{figure}[h] 
    \centering
    \includegraphics[width=0.75\linewidth]{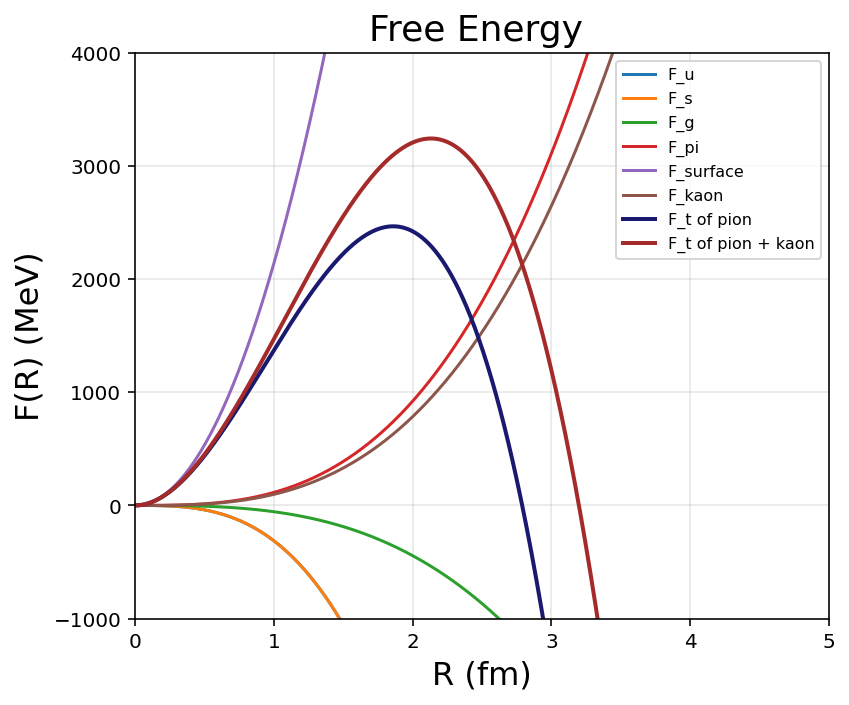}
    \caption{Comparison in variation of free energy in a kaonic+pionic vs a pionic medium when in a magnetic field and non-zero chemical potential}
    \label{fig:placeholder}
\end{figure}

\subsection{Effective Mass:}
 Fig. 3 graph shows the free energy of a system in the influence of a magnetic field and a non-zero chemical potential with different mass terms: rest mass, dynamic mass and effective mass.
\begin{figure}[h]
    \centering
    \includegraphics[width=0.65\linewidth]{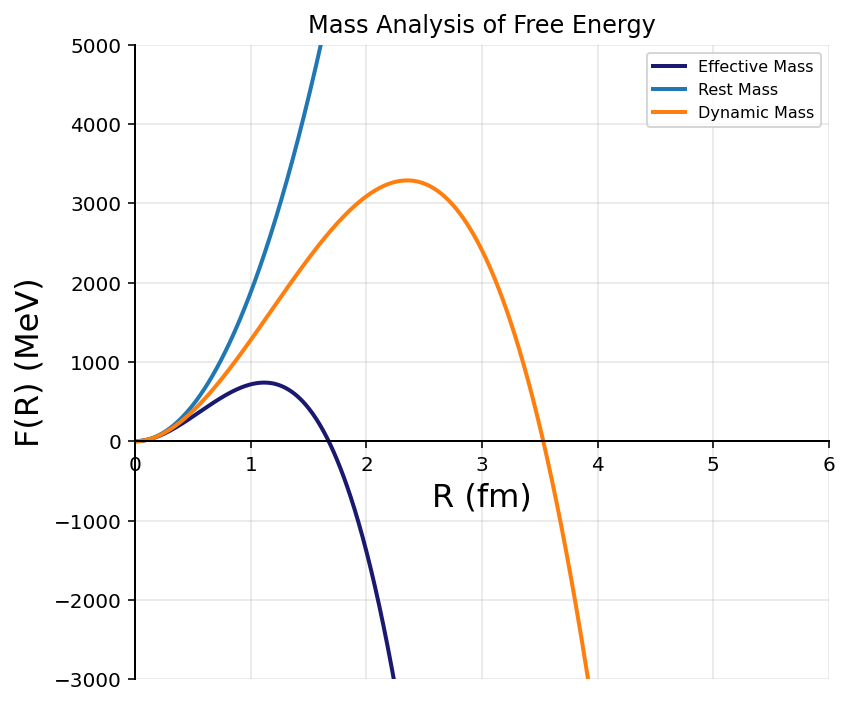}
    \caption{Comparison of variation in free energy when different mass terms are included}
    \label{fig:placeholder}
\end{figure}

It is observed that the system is unstable in the case of rest mass and there is no QGP droplet formation. In the case of the thermodynamic mass term, the free energy calculations reflect a bigger peak than the effective mass term, which includes the effect of both rest mass and thermodynamic mass. Hence, the system is less stable in the case of the real system in which rest mass and the thermodynamic effect on mass are both taken into account.

\subsection{Chemical Potential}

We see the effect of the chemical potential on the system in Fig. 4.
\begin{figure}[tb]
    \centering
    \includegraphics[width=0.6\linewidth]{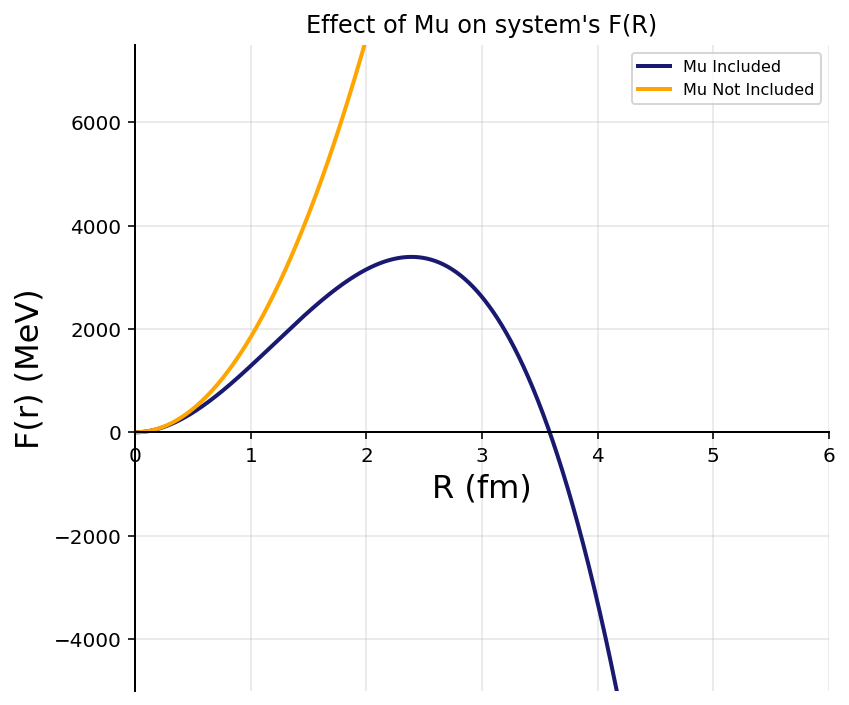}
    \caption{Comparison of variation in free energy when a zero and non-zero chemical potential are included}
    \label{fig:placeholder}
\end{figure}
It shows how including chemical potential in our calculations stabilises the system in the presence of a magnetic field. Hence, when the system is in the presence of an external magnetic field, a non-zero chemical potential is necessary for stability. We also verify the above by looking at the results of the free energy at different values of the chemical potential.

\begin{figure}[H]
    \centering
    \includegraphics[width=0.6\linewidth]{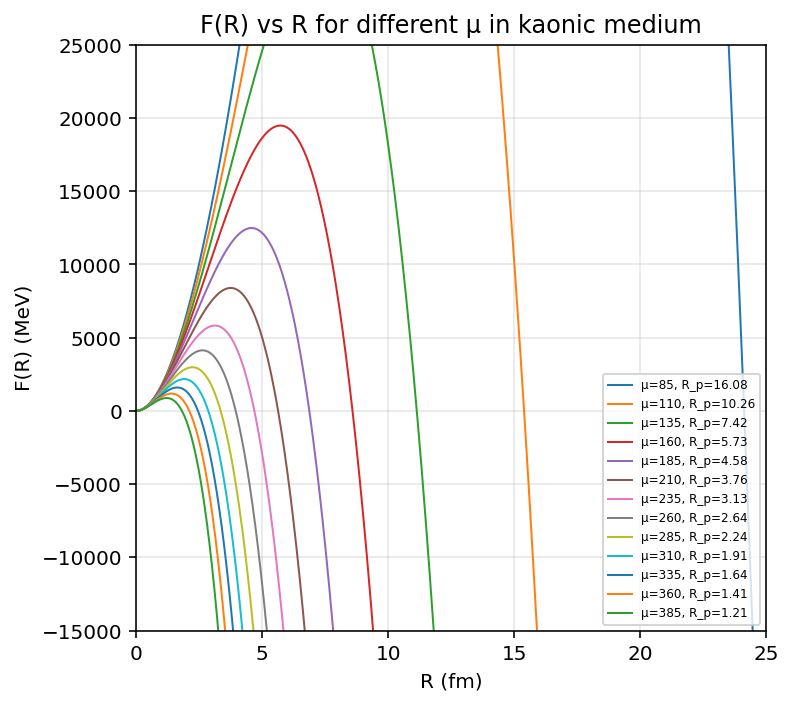}
    \caption{Variation in free energy as $\mu$ is changed along with corresponding radii values of the droplet}
    \label{fig:placeholder}
\end{figure}

These results show that the chemical potential and free energy, hence the stability of the system, have an inverse relationship. As the value of the chemical potential is increased, the peak of the free energy and the droplet radius decrease, leading to a less stable QGP.\\ 

To see the changing trend in free energy as a function of chemical potential, we study $F_{peak}$. We choose $F_{peak}$ for this analysis since it is the most characteristic value in the graph of free energy, since its value is directly related to the stability of the system. Hence, we plot a graph for values of $F_{peak}$ vs the corresponding values of $\mu$. We also check the same for the corresponding radii $R_{peak}$ of the droplet at $F_{peak}$.\\

\begin{figure}[h]
    \centering
    \includegraphics[width=1\linewidth]{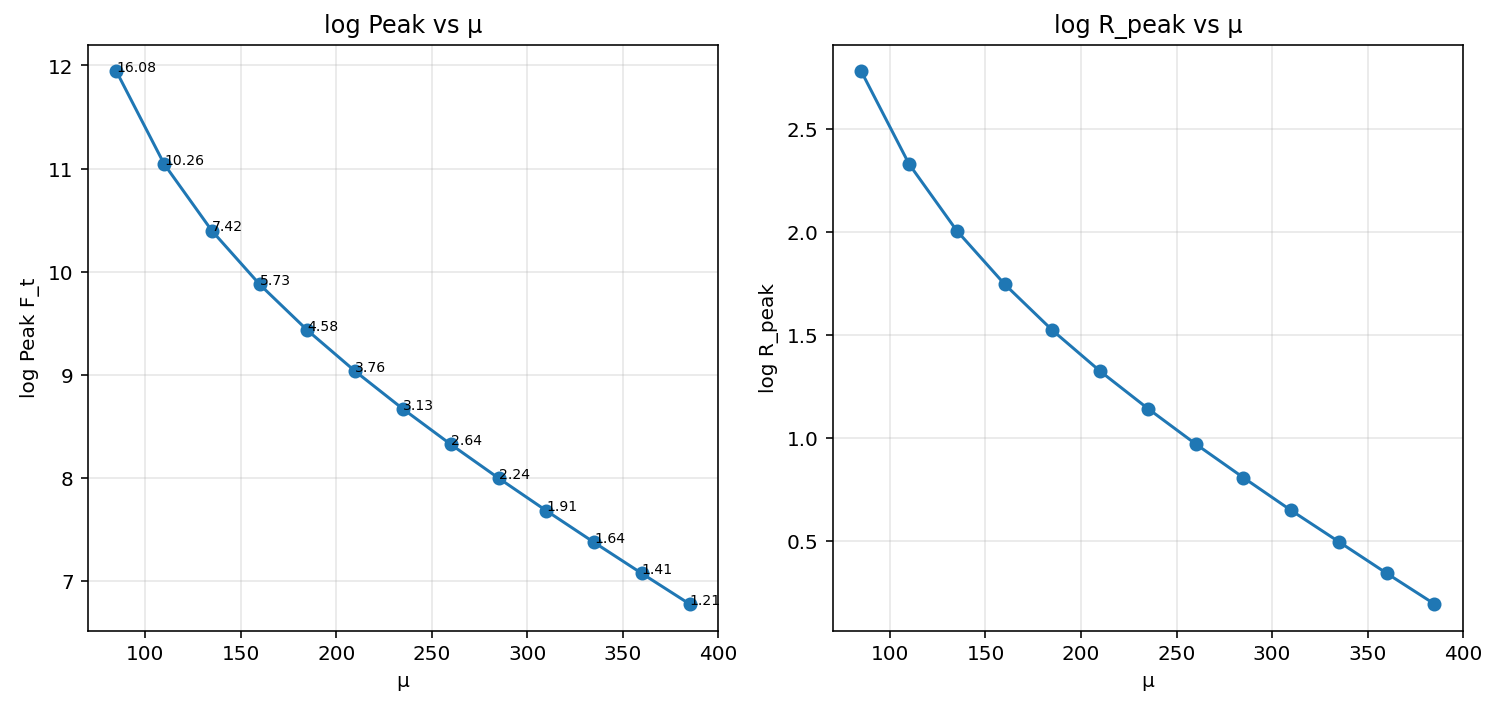}
    \caption{Variation in $\log F_{peak}$ and $\log R_{peak}$ as a function of $\mu$ showing a decreasing trend }
    \label{fig:placeholder}
\end{figure}

The results show us that the trend in free energy changes rapidly downwards as the chemical potential is increased. Due to this rapid change, we have plotted $\log F_{peak}$ and $\log R_{peak}$ vs $\mu$. \\

To observe the effect of kaon on the system, we have also plotted the same values, excluding kaon from the system. We observe that the kaonic system is just slightly shifted above the medium with only pions. Since this is a logarithmic graph, it implies that there is an exponential addition due to the inclusion of kaons in the system.   \\

\begin{figure}[tb]
    \centering
    \includegraphics[width=1\linewidth]{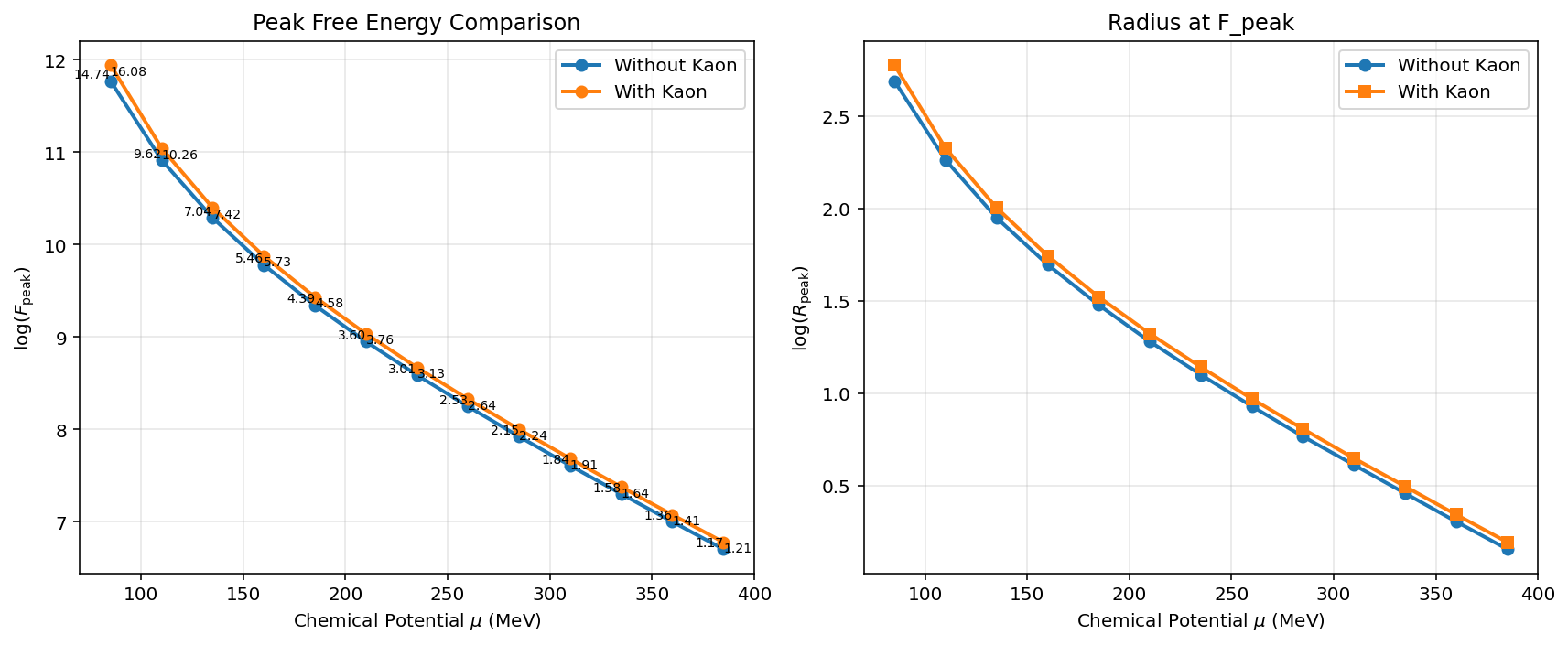}
    \caption{Comparing the change in $\log F_{peak}(\mu)$ and $\log R_{peak}(\mu)$ due to addition of kaon}
    \label{fig:placeholder}
\end{figure}

\subsection{Magnetic Field}

As a result of Lattice QCD calculations \cite{B.range}, we have expanded our magnetic field range to $eB=0\to45m_\pi^2$. We already know that an external magnetic field has a direct effect on the free energy, i.e. the stability of the QGP droplet\cite{magnetism}. To see the immediate effect of the magnetic field on the stability of the QGP droplet, we plot $F_{peak}$ and $R_{peak}$ as a function of multiples of $b$, where $b = 15m_\pi^2/e$; hence, our range extends from $0*b$ to $3*b$.
\begin{figure}[h]
    \centering
    \includegraphics[width=1\linewidth]{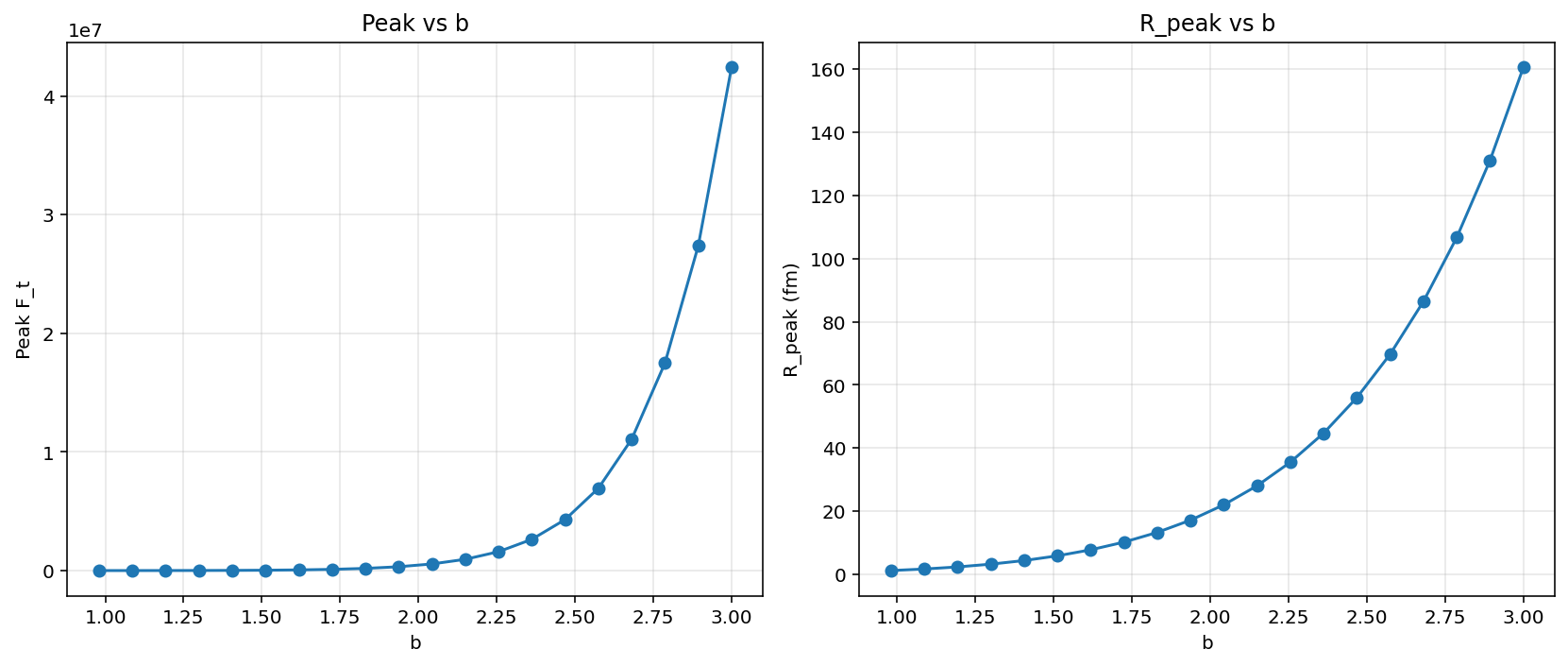}
    \caption{Variation in $F_{peak}$ and $R_{peak}$ as a function of magnetic field showing an increasing trend. }
    \label{fig:placeholder}
\end{figure}\\

The graph of $F_{peak}$ changes more slowly as a function of magnetic field when compared with the chemical potential graph. The same can be observed in the graph for $R_{peak}$.\\

We also analyse the effect of kaon in the system:\\

\begin{figure}[h]
    \centering
    \includegraphics[width=1\linewidth]{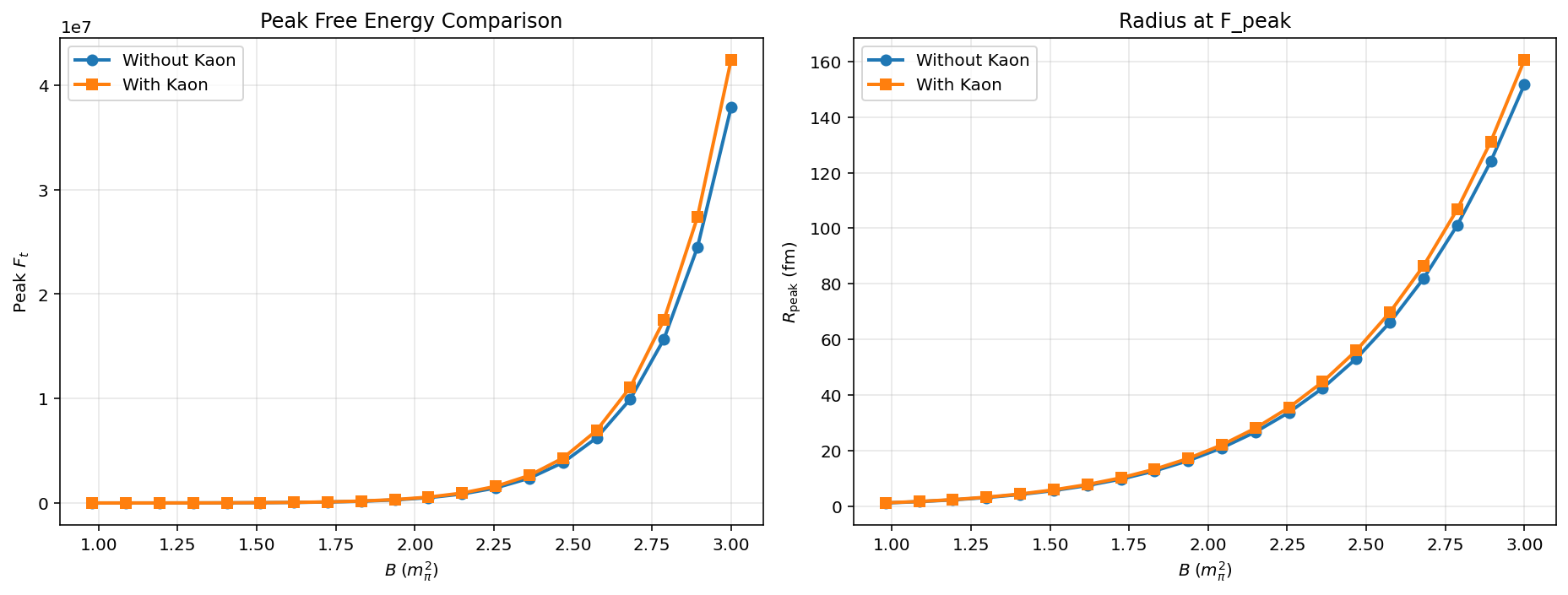}
    \caption{Comparing the change in $F_{peak}(\mu)$ and $R_{peak}(\mu)$ due to addition of kaon}
    \label{fig:placeholder}
\end{figure}
We observe that the effect of the magnetic field is similar in comparison to the chemical potential, in which there was a linear shift throughout in the logarithm of the values. We observe that the shift is non-linear, with the difference between $F_{peak}$ and $R_{peak}$ values increasing as the magnetic field increases; hence, the results diverge as the magnetic field gets stronger.

\subsection{$\mu$ and B Phase Diagrams}
In the above sections, we observe from the results of chemical potential and magnetic field that these two parameters have an inverse effect on the QGP droplet. Hence, we want to map out the behaviour of $F_{peak}$ for the entire range of these parameters and find the most stable configurations. This opposing nature of magnetic field and chemical potential thus allows us to contain a QGP droplet within a finite radius with a non-infinite peak in free energy. \\
\begin{figure}[H]
    \centering

    \begin{subfigure}{0.48\textwidth}
        \centering
        \includegraphics[width=\linewidth]{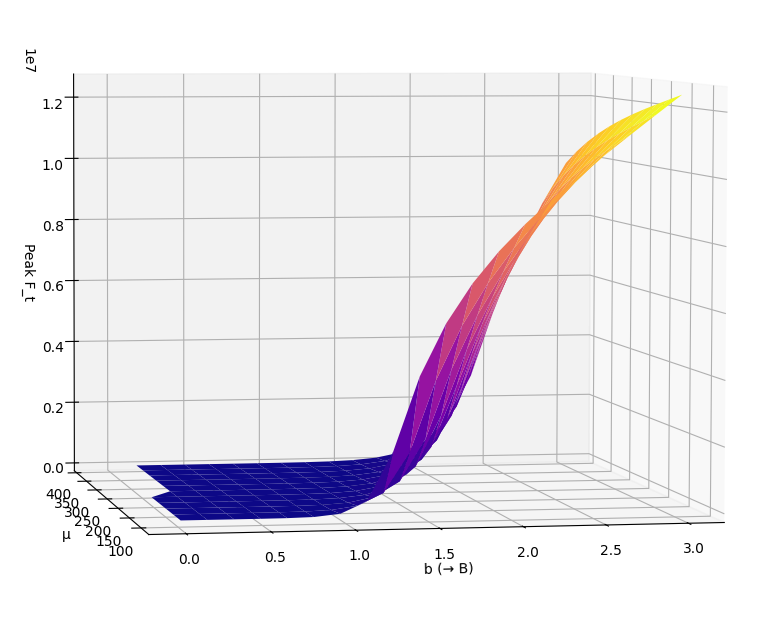}
        
        \label{fig:1}
    \end{subfigure}
    \hfill
    \begin{subfigure}{0.48\textwidth}
        \centering
        \includegraphics[width=\linewidth]{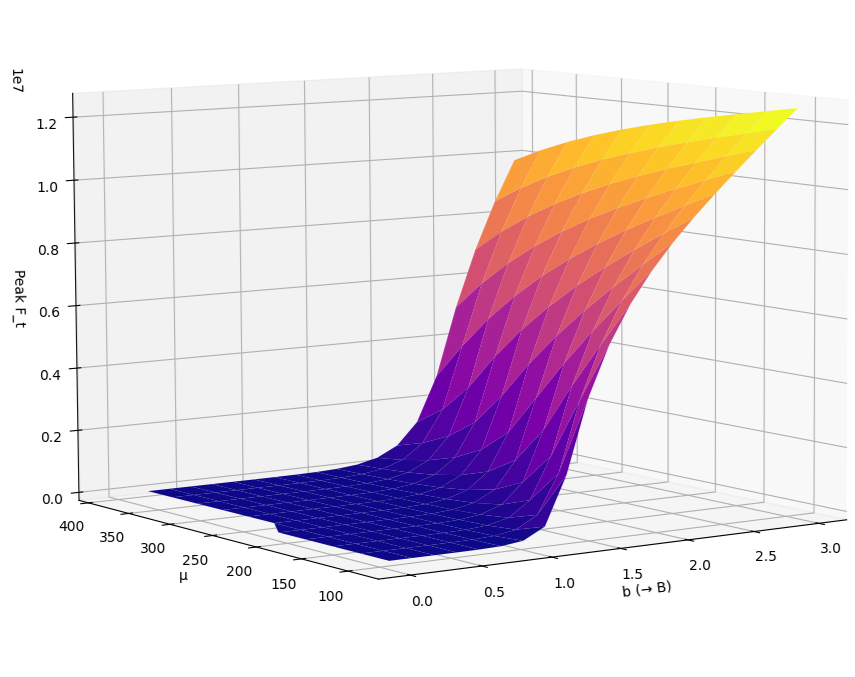}
        
        \label{fig:2}
    \end{subfigure}

    \vspace{0.5cm}

    \begin{subfigure}{0.48\textwidth}
        \centering
        \includegraphics[width=\linewidth]{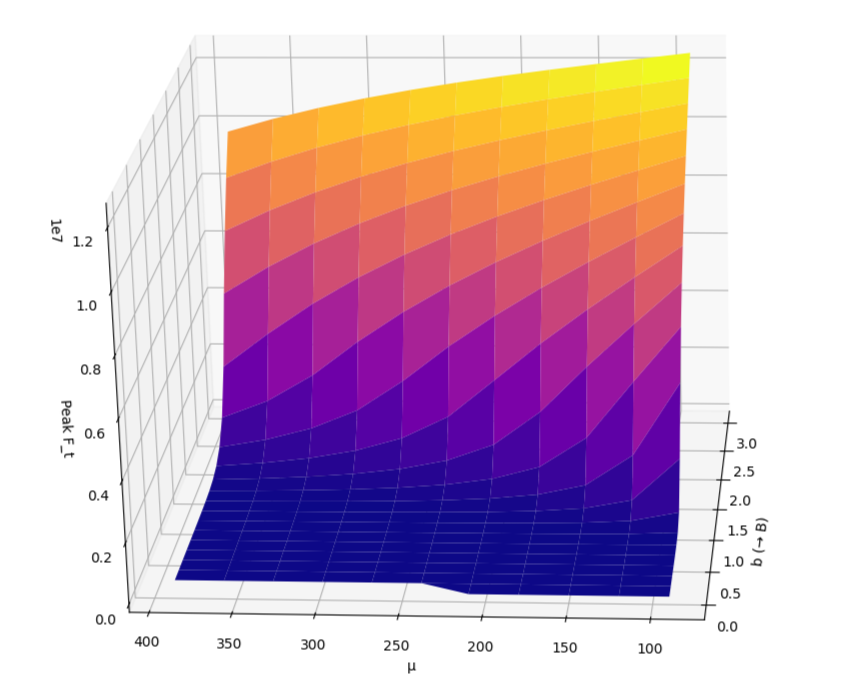}
        
        \label{fig:3}
    \end{subfigure}
    \hfill
    \begin{subfigure}{0.48\textwidth}
        \centering
        \includegraphics[width=\linewidth]{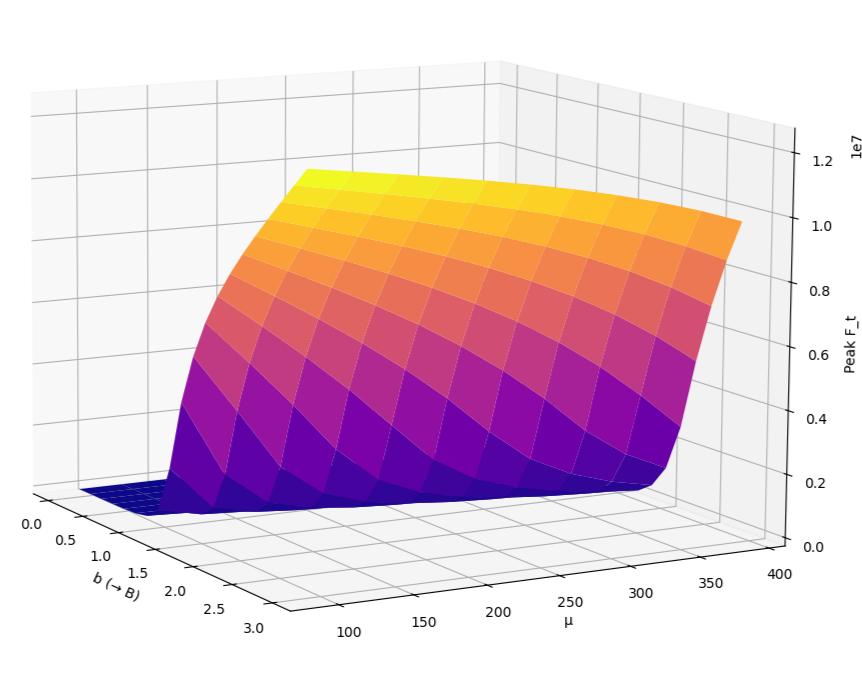}
        
        \label{fig:4}
    \end{subfigure}

    \caption{3D phase diagram between $\mu$ and B for $F_{peak}(\mu, B)$}
    \label{fig:fourplots}
\end{figure}

The phase diagram is plotted for the peak of the free energy as a function of chemical potential and magnetic field, $F_{peak}(\mu, B)$. Hence, by plotting a 3D graph for the same, we can directly see the trend of change in the stability of QGP droplet as both $\mu$ and B are changed. The plot increases non-linearly and plateaus at different rates across the values of $\mu$. The graph remains close to zero for initial values of $\mu$ and B, but takes off earlier for lower values of $\mu$ and starts rising later for higher values of $\mu$ as we move across values of $B$. The higher values of $F_{peak}$ are a virtue of the magnetic field. In them, we observe that for any value of B, the corresponding $F_{peak}$ value decreases as we move across $\mu$, which should be obvious given our preceding discussion. This leads to a non-linear increasing graph across the values of the highest free energies. This phase diagram, hence, gives us a collective image of the change in stability of the system as a function of parameters of chemical potential and magnetic field in the entirety of the spectrum of their values.

The heat map for the phase diagrams above looks like the following:
\begin{figure}[H]
    \centering
    \includegraphics[width=0.75\linewidth]{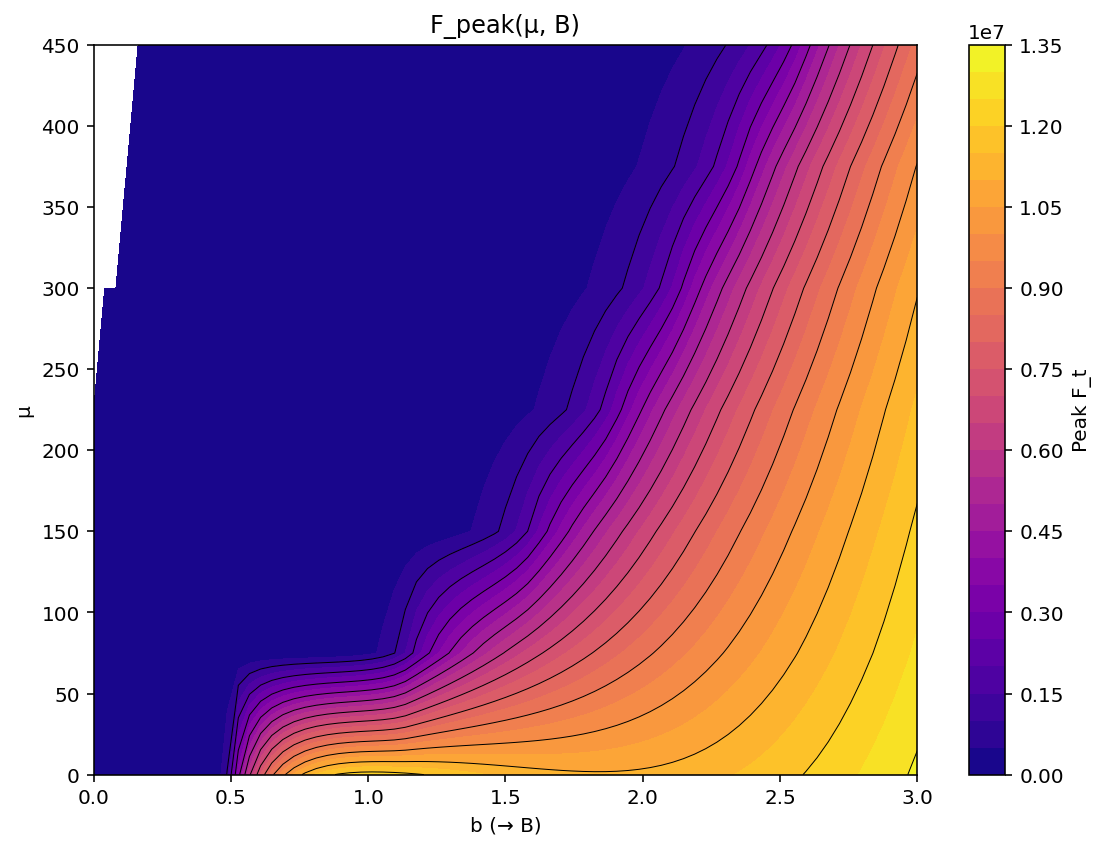}
    \caption{Heat Map for $F_{peak}(\mu,B)$}
    \label{fig:placeholder}
\end{figure}
 The above analysis for the phase diagrams becomes clearer through heat maps. The free energy increases faster for lower values of $\mu$ and slower for larger values of $\mu$ as the strength of the external magnetic field is increased. We observe that as the magnetic field is increased, we get higher free energy values, albeit non-linearly, due to the chemical potential. The contour lines show the trend in the change of free energy. The path followed by contour lines represents how a value or a set of values moves across the parameter space. Hence, in this heat map we observe that the trend for free energy peak values is a non-linear curve that increases, decreases and then has an increasing trend. The most stable values can be found for high values of the magnetic field and low values of the chemical potential. As the chemical potential is increased, we need to provide the system with higher magnitudes of magnetic field to form a stable QGP droplet.\\
 
 The blank area in the top left corner represents areas where the equations of our system do not allow for the formation of QGP droplet. Those configurations of low magnetic field and high chemical potential cannot lead to the formation of a QGP droplet, let alone of high or low stability.\\

  To also observe the effect of the kaon across the various configurations of magnetic field and chemical potential, the below heat map was produced. Not much difference is observed in the heat map due to kaon. The most noticeable change lies in the middle range of the magnetic field and the lower range of the chemical potential. In that region, a slightly more stable QGP can be found in the kaonic + pionic medium system.
  
\begin{figure}[H]
    \centering
    \includegraphics[width=0.75\linewidth]{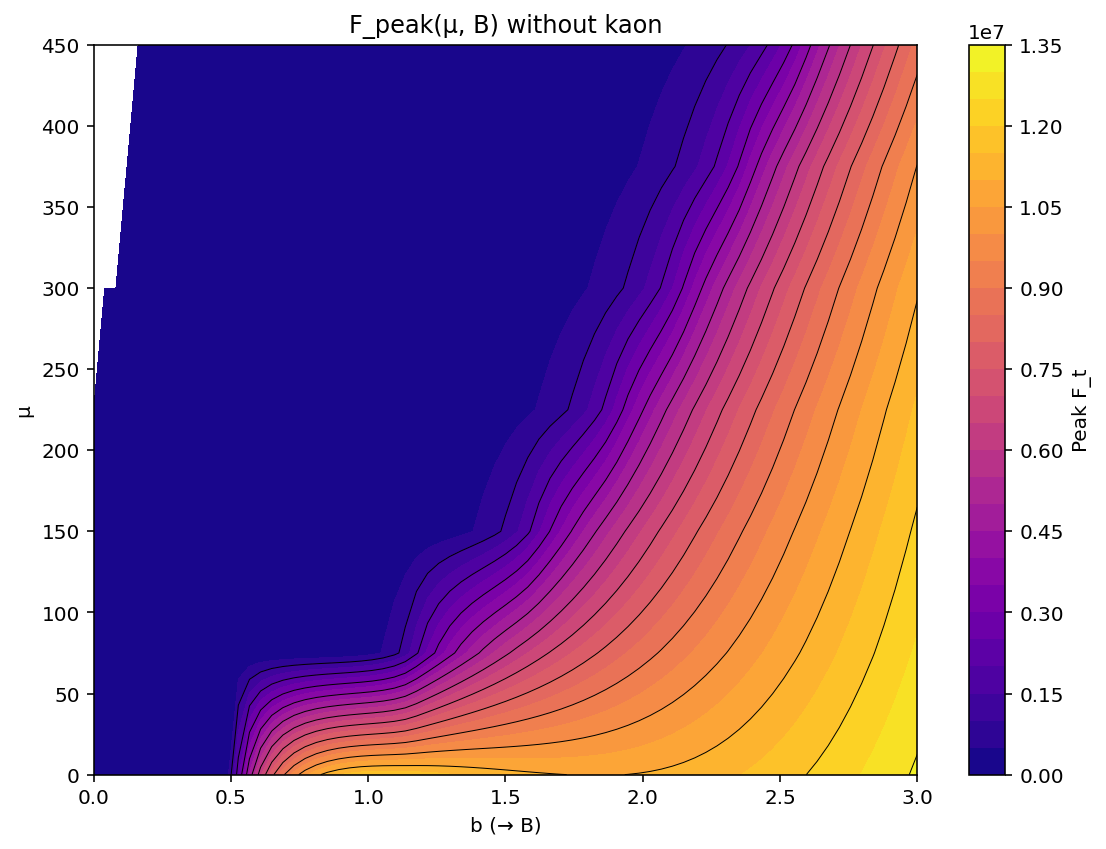}
    \caption{Heat Map for $F_{peak}(\mu,B)$ without kaonic medium}
    \label{fig:placeholder}
\end{figure}

\subsection{Entropy and Heat Capacity}

  In the previous study, which included an external magnetic field, non-zero chemical potential and a dynamic mass term, a discontinuity was observed in the graph for entropy at the critical transition temperature $T_C=175$ MeV\cite{magnetism}. This discontinuity in the graph for entropy points towards a first-order phase transition. We investigate if the system still undergoes a first-order phase transition after other parameters are added, such as kaons in the medium of the QGP droplet and an effective mass term.\\
  
\begin{figure}[h]
    \centering
    \includegraphics[width=0.7\linewidth]{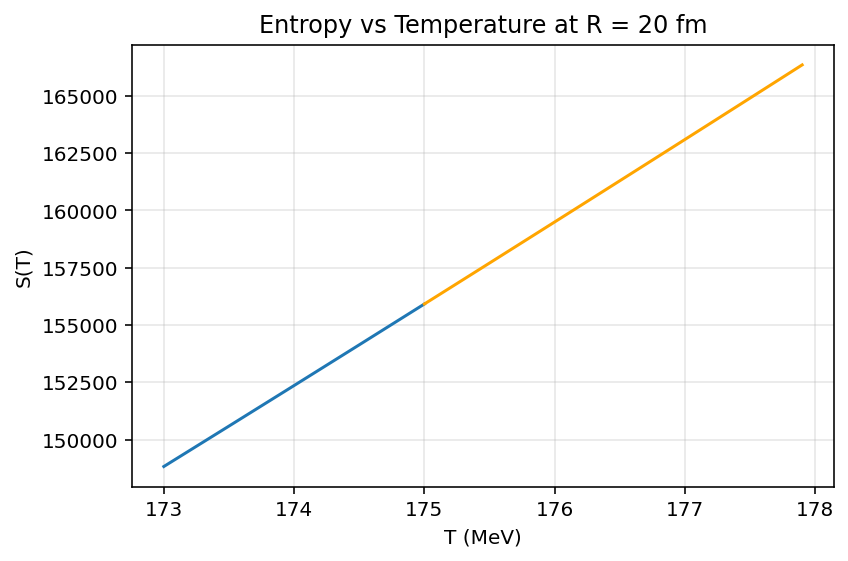}
    \caption{Entropy vs Temperature graph showing no discontinuity at $T_C$=175MeV}
    \label{fig:placeholder}
\end{figure}

  No discontinuity was observed in the graph at the critical transition temperature $T_C=175$ MeV. Hence, upon these additions, the system no longer goes through a first-order transition. For more information about the phase transition, we look at the graph for heat capacity vs temperature.
  
\begin{figure}[h]
    \centering
    \includegraphics[width=0.75\linewidth]{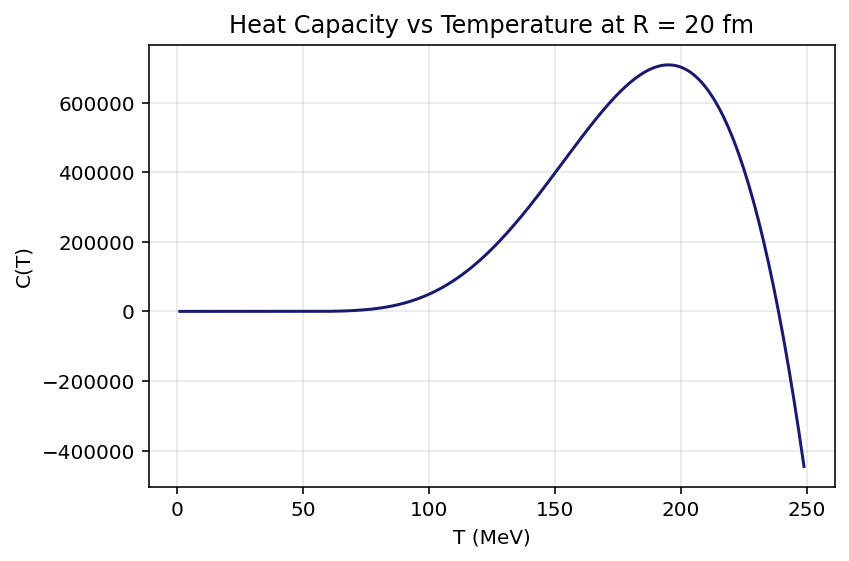}
    \caption{Heat Capacity vs Temperature graph}
    \label{fig:placeholder}
\end{figure}
The result for heat capacity differs greatly from the previous study, in which an ever-increasing graph was shown \cite{magnetism}. Our graph here increases exponentially and then has a downward trend. This decreasing trend points towards a phase transition. Hence, it may be concluded that the system undergoes a weakly first-order phase transition.